\documentclass[apjl]{emulateapj}
\usepackage{apjfonts}

\begin{document}
\submitted{Accepted for publication in the Astrophysical Journal Letters}
\title{Hide-and-Seek with the Fundamental Metallicity Relation}
\author{
D.~Kashino\altaffilmark{1}, 
A.~Renzini\altaffilmark{2,3},
J.~D.~Silverman\altaffilmark{4},
E.~Daddi\altaffilmark{5}
}

\altaffiltext{1}{
Division of Particle and Astrophysical Science, Graduate School of Science, Nagoya University, Nagoya, 464-8602, Japan\\
\email{daichi@nagoya-u.jp}
}
\altaffiltext{2}{
INAF Osservatorio Astronomico di Padova, vicolo dell'Osservatorio 5, I-35122 Padova, Italy
}
\altaffiltext{3}{
National Astronomical Observatory of Japan, Mitaka, Tokyo 181-8588, Japan
}
\altaffiltext{4}{
Kavli Institute for the Physics and Mathematics of the Universe (WPI), Todai Institutes for Advanced Study, the University of Tokyo, Kashiwanoha, Kashiwa, 277-8583, Japan
}

\altaffiltext{5}{
Laboratoire AIM-Paris-Saclay, CEA/DSM-CNRS-Universit{\' e} Paris Diderot, Irfu/CEA-Saclay, Service d'Astrophysique, F-91191 Gif-sur-Yvette, France
}

\begin{abstract}
We use $\sim 83,000$  star-forming galaxies at $0.04<z<0.3$ from the Sloan Digital Sky Survey to study the so-called fundamental metallicity relation (FMR) and report on the disappearance of its anti-correlation between metallicity and star formation rate (SFR) when using the new metallicity indicator recently proposed by Dopita et al.  In this calibration, metallicity is primarily sensitive to the  emission line ratio [N{\sc ii}]$\lambda 6584$ / [S{\sc ii}]$\lambda \lambda$ 6717, 6731 that is insensitive to dilution by pristine infalling gas that may drive the FMR anti-correlation with SFR. Therefore, we conclude that the apparent disappearance of the FMR (using this new metallicity indicator) does not rule out its existence.

\end{abstract}

\newpage

\section{Introduction}

An anti-correlation of the gas-phase metallicity ($Z$) and star formation rate (SFR) at a fixed stellar mass ($M_\ast$) has been first reported by \cite{2008ApJ...672L.107E} and then further explored and discussed in many subsequent studies (e.g., \citealt{2010A&A...521L..53L,2013MNRAS.434..451L,2011MNRAS.414.1263M,2012MNRAS.422..215Y,2013ApJ...765..140A,2014ApJ...789L..40W,2014ApJ...792...75Z}).  In particular, \citet[][hereafter M10]{2010MNRAS.408.2115M} proposed the so-called fundamental metallicity relation (FMR),  as a redshift-invariant surface in the ($M_\ast$, SFR, $Z$), with SFR being a third axis in the mass--metallicity (MZ) relation.  Such an FMR was then meant to describe both anti-correlated SFR and $Z$ fluctuations at fixed mass and the redshift evolution of metallicity and SFR, where, with increasing lookback time, the specific SFR (sSFR) goes up and $Z$ goes down, thus keeping galaxies on the FMR.

However, large discrepancies still exist concerning the size and shape of the SFR$-Z$ anti-correlation in the literature. For example, the anti-correlation is noticeable in M10 at low masses but nearly vanishes at high masses, whereas in \citet{2013ApJ...765..140A} it is nearly equally strong at all masses and much stronger than in M10. On the contrary, \citet{2012MNRAS.422..215Y} shows that the metallicity increases, rather than decreases, with increasing SFR for massive galaxies. Moreover, it is not clear whether an sSFR$-Z$ anti-correlation exists at all at high redshift (e.g., \citealt{2014ApJ...795..165S,2014ApJ...789L..40W,2014ApJ...792...75Z,2016ApJ...822..103G}), or whether high redshift galaxies follow the FMR proposed by M10.  These discrepancies may largely arise from the use of different metallicity indicators or different galaxy selection criteria adopted in different studies.

In a physical interpretation of the FMR, upward fluctuations in the amount of pristine infalling gas would boost star formation while diluting the metal abundance of the ISM \citep{2008ApJ...672L.107E,2010MNRAS.408.2115M}.  \citet{2013ApJ...772..119L} have introduced a physically motivated model that predicts the metallicity of the ISM as a function of $M_\ast$ and SFR, with infalling and outflowing gas regulating star formation and chemical enrichment in a galaxy \citep[see also][]{2013MNRAS.430.2891D}.  Such an idea has also been suggested from cosmological hydrodynamical simulations \citep{2012MNRAS.421...98D}.  This model unifies, using a simple relation, both the local up and down fluctuations of sSFR and metallicity as well as their secular evolution with redshift; see also \citet{2014ApJ...792....3M}. On physical grounds, one expects an FMR to exist, but observations are still somewhat contradictory as to whether an FMR actually exists, and if so, what is its shape is at low and high redshifts.

Recently, a new metallicity calibration has been proposed by \citet[][hereafter D16]{2016Ap&SS.361...61D} that differs substantially from previous ones.  Oxygen is an $\alpha$-capture primary element  produced by massive stars, whereas nitrogen has both a primary and secondary component (coming from the conversion of carbon and oxygen originally present in stars) and is produced both by short- (massive) and long-leaving (intermediate-mass) stars \citep[e.g.,][]{1981A&A....94..175R,2016MNRAS.458.3466V}. Therefore, the N/O ratio increases with increasing metal abundance of the ISM \citep[e.g.,][]{2002ApJS..142...35K} and with an increasing contribution by intermediate-mass stars, hence, on the specific star formation history of individual galaxies.  The novelty of the D16 calibration is the use of the line ratio [N{\sc ii}]/[S{\sc ii}] as a proxy for the N/O ratio, where sulfur, like oxygen, is an $\alpha$-capture {\it primary} element.  The original FMR, as in M10, was based on the traditional indicators, the [N{\sc ii}]/H$\alpha$ and ([O{\sc ii}]$+$[O{\sc iii}])/H$\beta$ ratios, following the calibration as in \citet[][hereafter M08]{2008A&A...488..463M}. 

In this paper, we investigate the $Z$--$M_\ast$--SFR (i.e., FMR) relation while comparing the D16 and M10 calibrations for determining the gas-phase metallicity. Throughout this Letter, we use a \citet{2003PASP..115..763C} initial mass function (IMF).

\section{Sample}

Our galaxy sample is extracted from the Sloan Digital Sky Survey (SDSS) Data Release 7 \citep{2009ApJS..182..543A}, while the physical quantities of galaxies are based on the MPA-JHU catalog \citep{2003MNRAS.341...33K,2004MNRAS.351.1151B,2004ApJ...613..898T} of Data Release 12 \citep{2015ApJS..219...12A}, which provides the total SFRs.  SFRs are derived from the H$\alpha$ luminosity, for which dust extinction and fiber aperture losses are corrected, for star-forming galaxies in our sample \citep{2004MNRAS.351.1151B}.  We use stellar masses derived using Le Phare \citep{2011ascl.soft08009A} and taking into account the emission lines in the spectral energy distribution (SED) fitting (see \citealt{2011ApJ...730..137Z} for details).  SFRs in the MPA-JHU catalog are converted to a Chabrier IMF by subtracting 0.05~dex from the original values.  We note that M10 used SFRs measured within the fiber aperture, which are smaller (by roughly one-half) than the total SFRs used in this study.  We present here results based on the total (aperture-corrected) SFRs, but we emphasize that the same results and conclusions are reached when using in-fiber SFRs.

Galaxies are selected over a redshift range of $0.04<z<0.3$ to ensure that the [S{\sc ii}] doublet lines fall within the wavelength range of the SDSS spectrograph (3800--9200 \AA).  The lower redshift limit is imposed to reduce the aperture effects.  According to \citet{2005PASP..117..227K}, the line measurements of galaxies at $z<0.04$ tend to be highly biased toward the central area, which is typically more enriched, as the covering fraction of the SDSS fiber is typically less than $\sim 20\%$.  We stress  that the upper limit of the redshift range  is the same as in M10 and that our conclusions do not change when the sample is restricted to a narrower redshift range (e.g., $0.04<z<0.1$), to reduce the effects of the sSFR and $Z$ evolution with redshift. 

We restrict the sample to galaxies having emission-lines detections of H$\alpha$ with $\textrm{S/N}>5$ and [N{\sc ii}]$\lambda$6584, [S{\sc ii}]$\lambda \lambda$6717, 6731, H$\beta$, [O{\sc iii}]$\lambda$5007, and [O{\sc ii}]$\lambda \lambda$3726,3729 with $\textrm{S/N}>3$.  Following \citet{2006MNRAS.371..972S}, we distinguish {\it star-forming} galaxies from AGNs  by applying  the simplified formulation derived by \citet{2010MNRAS.403.1036C}:
\begin{eqnarray}
&& N2 \equiv \log \left( [\textrm{N{\sc ii}}]\lambda 6584 / \mathrm{H\alpha}\right) < -0.2, \\
&& \log \left( [\textrm{O{\sc iii}}]\lambda 5007/\mathrm{H\beta} \right) < \frac{0.29}{N2+0.20}+0.96.
\end{eqnarray}
Our selection criteria differs from that of M10, who adopted $\textrm{S/N}>25$ for only H$\alpha$ without imposing limits on the other lines. We have checked that our results and conclusions do not change when using the same criterion adopted in M10.

\section{Metallicity determination}
\label{sec:D16}

For a proper comparison with the result of M10, we further impose a selection based on the method of metallicity determination.  For the selected star-forming galaxies, gas-phase oxygen abundances are estimated using two independent indicators, the [N{\sc ii}]/H$\alpha$ ratio and the $R_{23}$ index, defined as $R_\mathrm{23} = \log \left([\textrm{O{\sc ii}}]\lambda\lambda3726,3729 + [\textrm{O{\sc iii}}]\lambda\lambda4959, 5007 \right)/\mathrm{H\beta}$.  We correct the $R_{23}$ values for dust extinction based on the Balmer decrement (H$\alpha$/H$\beta$) assuming a \citet{2000ApJ...533..682C} extinction curve.  As in M10, we adopt  the M08 calibrations and select only galaxies in which the two estimates agree within 0.25~dex (98\% of the sample), for which the metallicity is defined as an average of the two estimates.  Note that our conclusions do not depend on whether this criterion is applied or not.  The resulting sample consists of 83,076 galaxies.

For the same galaxies, the metallicity was determined also following the D16 calibration using the line ratios [N{\sc ii}]/[S{\sc ii}] and [N{\sc ii}]/H$\alpha$, which  is given to hold over the metallicity range of $8<12+\log (\mathrm{O/H})<9$:
\begin{equation}
12+\log (\mathrm{O/H}) = 8.77 + N2S2 +0.264 N2
\label{eq:D16}
\end{equation}
where $N2S2=\log ([\textrm{N{\sc ii}}]\lambda 6584 / [\textrm{S{\sc ii}}]\lambda \lambda 6717, 6731)$ and $N2 = \log ([\textrm{N{\sc ii}}]\lambda 6584 / \mathrm{H\alpha})$.  
This calibration has the advantage that it is almost independent of the ionization parameter and gas pressure (see Figure 2 in D16), as both the [N{\sc ii}] and [S{\sc ii}] lines respond to the changes in the ionization parameter and/or gas pressure in a similar manner, and the wavelength proximity of the lines ensures little dependence on reddening.  

This relation is based  on the assumption N/S $\simeq$ N/O and on the N/O vs. O/H relation derived from local calibrators (Figure 1 in D16), where N/O increases with increasing metallicity as the secondary nitrogen production becomes predominant.  We also note that variations of the S/O abundance ratio are expected to be  small ($\sim 0.1$~dex), and indeed the S/O ratio is almost independent of metallicity \citep{2006A&A...448..955I,2009MNRAS.398..949P}.  Thus, there is ample support for using  the [N{\sc ii}]/[S{\sc ii}] ratio as a proxy for the N/O ratio.  However, as mentioned by D16, it remains to be demonstrated whether the N/O vs. O/H relation  implicit in Equation \ref{eq:D16} holds also at high redshifts. This is to say that the metallicity scale given by Equation \ref{eq:D16} must hold for the galaxies used to calibrate it, or having had similar chemical evolutionary histories, and may not apply to galaxies having experienced a different chemical evolution compared to local calibrators, which may well include all high redshift galaxies.

\begin{figure}[tbp]
   \centering
   \includegraphics[width=3.5in]{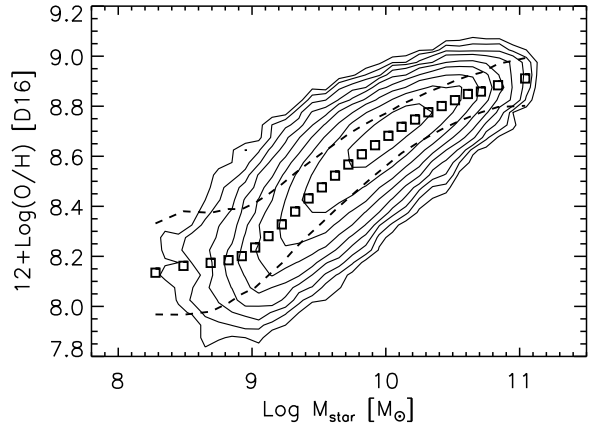} 
   \caption{Stellar mass--metallicity relation for our sample of 83,076 galaxies ($0.04<z<0.3$) where metallicity is derived from  Equation \ref{eq:D16}.  Contours show the number distribution of galaxies  in log scale.  Squares indicate the median metallicity in bins of stellar mass and the dashed lines indicate the central 68th percentile of each bin.}
   \label{fig:MZ}
\end{figure}

\section{Results}

\begin{figure*}[tbp]
   \centering
   \includegraphics[width=5in]{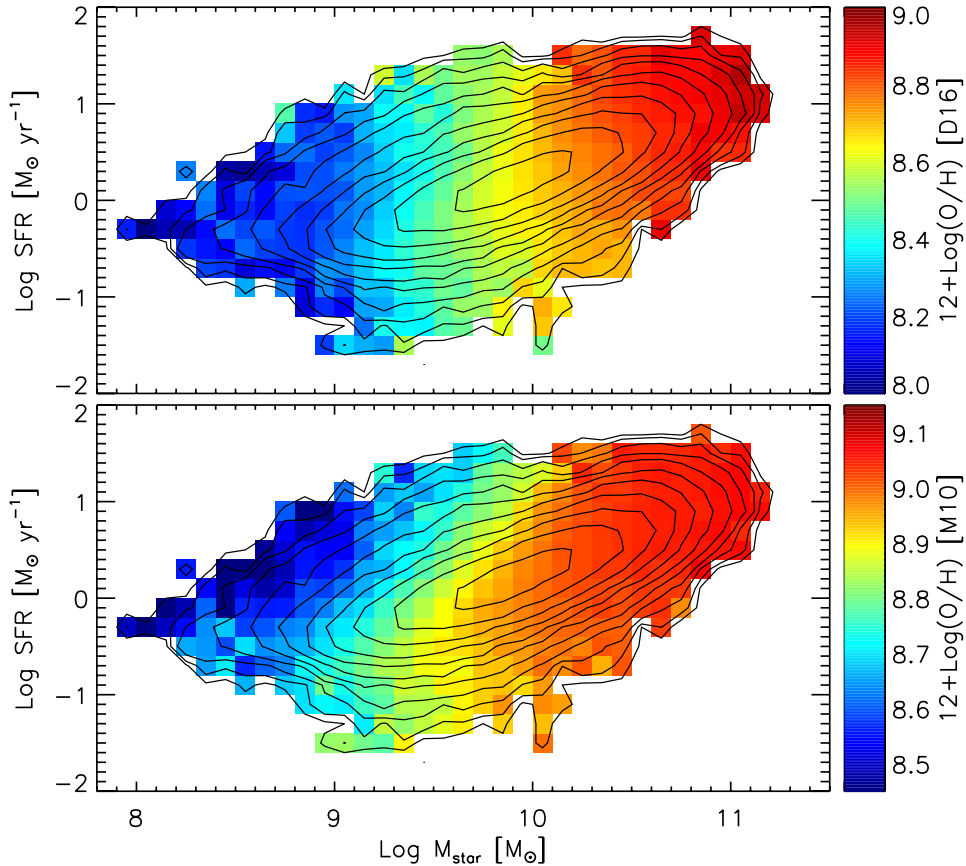} 
   \caption{The SFR$M_*$ relation color-coded with metallicity derived from  the D16 calibration (Equation \ref{eq:D16}; upper panel) or like in the \citet{2010MNRAS.408.2115M} determination using the $N2$ and $R_\mathrm{23}$ indices (bottom panel).  Contours indicate the number count in cells in log scale.}
   \label{fig:MxSFR}
\end{figure*}

Figure \ref{fig:MZ} shows the MZ relation based on the D16 calibration.  At $M_\ast\gtrsim10^9~M_\odot$, metallicity is strongly correlated with stellar mass, as shown in the MZ relations based on other indicators such as the [N{\sc ii}]/H$\alpha$ ratio (e.g., \citealt{2014ApJ...792...75Z}).  We find the scatter $\sigma (\log (\mathrm{O/H})) = 0.12~\mathrm{dex}$ ($8<\log M_\ast /M_\odot < 11.3$), which is small relative to the entire metallicity range spanned by the sample ($\sim 1~\mathrm{dex}$) and similar to that found  in previous studies with other indicators ($\sim 0.1~\mathrm{dex}$; e.g., \citealt{2004ApJ...613..898T}).  This tight correlation strongly supports a notion according to which, in more massive galaxies, the chemical enrichment has proceeded to a more advanced level, hence, enhancing the secondary-to-primary element ratio.  Interestingly, it is seen that the median metallicity is almost constant at low masses (below $10^9~M_\odot$).  In such less massive  galaxies, the primary nitrogen production may be dominant; thus, the secondary-to-primary element ratio (i.e., N/O and N/S) is almost constant at a value that is determined by the physics of the primary nucleosynthesis in massive stars ($N2S2\sim -1.5$, e.g., \citealt{2014ApJ...785..153M}). 
So, the flattening is due to $N2S2$ in Equation \ref{eq:D16} approaching this pedestal level. This trend is not seen in the MZ relations that have been reported based on the strong line methods (e.g., see Figure 6 of \citealt{2004ApJ...613..898T}), where the metallicity continuously declines with decreasing stellar mass.  

Figure \ref{fig:MxSFR} shows the SFR$-M_\ast$ relation, color-coded by metallicity as from the D16 calibration, Equation \ref{eq:D16}, (upper panel) or based on the M10 determination using $N2$ and $R_{23}$ indices (bottom panel).  Galaxies are separated in ($M_\ast$, SFR) bins with $\Delta \log M_\ast=0.1$~dex and $\Delta \log \mathrm{SFR}=0.2$~dex.  The cells including five galaxies or more are color-coded to indicate the median metallicity.  In both panels, the median metallicity increases with increasing stellar mass.  However, the dependence on the SFR is obviously different between the two metallicity calibrations.  In the bottom panel, the FMR is clearly visible, showing metallicity decreasing with increasing SFR at a fixed stellar mass.  In particular, the anti-correlation is clearly seen at $10^{8.5} \lesssim M_\ast/M_\odot \lesssim10^{10}$.  In contrast, the D16 metallicity is nearly invariant with the SFR at a fixed stellar mass over the entire stellar mass range probed by the sample.

\begin{figure}[htbp]
   \centering
   \includegraphics[width=3.5in]{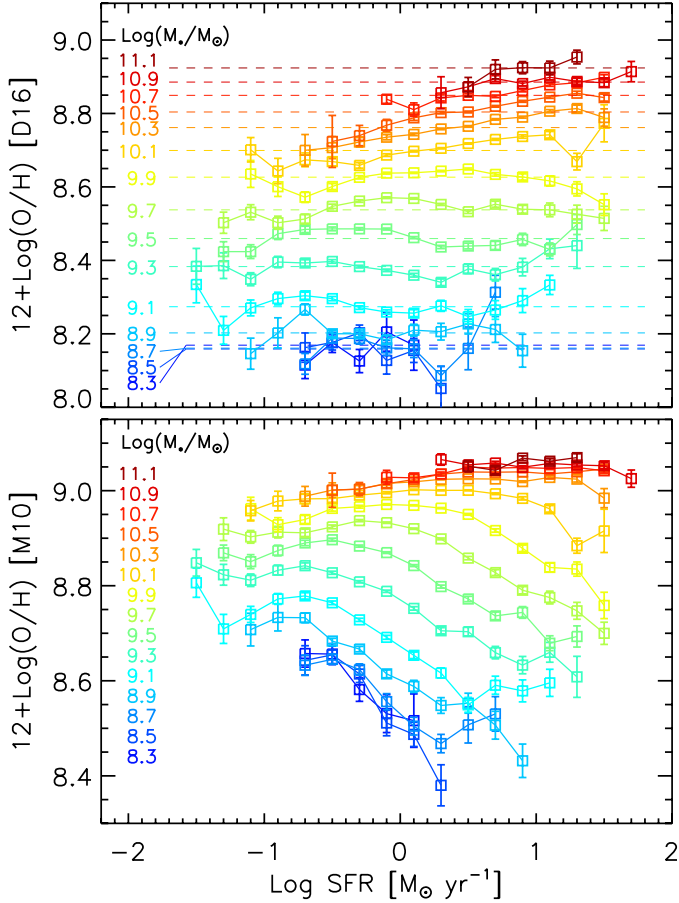}
   \caption{Median metallicity  computed with either Equation \ref{eq:D16} (upper panel) or the M10 calibration (bottom panel), as a function of SFR.  Squares indicate median metallicities in each ($M_\ast$, SFR) cell ($\Delta \log M_\ast=0.2$~dex, $\log \mathrm{SFR}=0.2$~dex).  Error bars indicate $\sigma / \sqrt{N}$ where $\sigma$ is a standard deviation and $N$ is the number of galaxies in each ($M_\ast$, SFR) bin.  Horizontal dashed lines (top panel) indicate the median metallicities measured in each stellar mass bin.}
   \label{fig:ZxSFR}
\end{figure}

The disappearance of the SFR dependence of metallicity is even more clearly seen in Figure \ref{fig:ZxSFR}, showing the median metallicities with both calibrations as a function of SFR for galaxies with  different masses as labelled.  Galaxies are binned by $\Delta \log \mathrm{SFR}=0.2$~dex and $\Delta \log M_\ast=0.2$~dex.  The data points indicate the value for the cell including 10 galaxies or more.  The bottom panel shows that the $N2+R_{23}$-based metallicity decreases with increasing SFR.  The change in  metallicity tends to be larger for less massive galaxies, while the metallicity is almost invariant with SFR at $M_\ast \gtrsim 10^{10}~M_\odot$, as shown in Figure \ref{fig:MxSFR}.  In contrast, the top panel of Figure \ref{fig:ZxSFR} clearly shows that the D16 metallicity is almost independent of SFR at a fixed stellar mass.

Thus, we recover the FMR anti-correlation as in M10 when we use  metallicities as derived in  M10, but the anti-correlation, hence the FMR, apparently disappear when using the D16 metallicity scale. So, does the FMR really exist, or is it a mere artifact of a specific metallicity scale?

We also notice from Figure \ref{fig:ZxSFR} (top panel) that metallicity increases slightly with increasing SFR in galaxies with $M_\ast \gtrsim 10^{10}~M_\odot$.  This {\it positive} correlation between $Z$ and SFR is statistically significant, while being opposite to the original FMR.  Such an {\it inverse} trend at high masses has also been pointed out also by \citet{2012MNRAS.422..215Y}, and Figure  \ref{fig:ZxSFR} indicates that it becomes  stronger when using the D16 metallicities, while present also when using the M08 calibration. Clearly, dilution with pristine gas cannot be invoked to account for this trend.

\section{Conclusions}

We measure the metallicity of star-forming galaxies based on two distinct calibrations:  one recently proposed by \citet{2016Ap&SS.361...61D} incorporating both the [N{\sc ii}]/[S{\sc ii}] line ratio and $N2$ index and the traditional one based on the $N2$ and $R_{23}$ indices (as calibrated by \citealt{2008A&A...488..463M}) using $\sim 83,000$ local star-forming galaxies ($0.04<z<0.3$) from the SDSS. We find that an FMR (i.e., a $Z$--SFR anti-correlation at a fixed stellar mass) exists or does not exist depending on how metallicity is measured.   In particular, the SFR dependence in the MZ relation disappears when using the D16 metallicity, while we retrieve it when using the same metallicity scale adopted by M10.  

However, ironically enough, we maintain that the disappearance of the FMR with D16 is actually consistent with its existence.  Indeed, if enhancement of star formation and dilution of gas-phase oxygen abundance are both driven by infall of pristine/metal-poor gas, one does not expect the N/S and N/O ratios to change at all. Now, in Equation \ref{eq:D16}, the oxygen abundance depends linearly on the [N{\sc ii}]/[S{\sc ii}] ratio (which does not change by dilution) and only to the $\sim 1/4$ power of the [N{\sc ii}]/H$\alpha$ ratio, whereas both ratios span a similar range ($\sim 1$ dex). Hence, being primarily sensitive to the 
[N{\sc ii}]/[S{\sc ii}] ratio, the D16 metallicity  is almost insensitive to dilution effects that may be present in low as well as high redshift galaxies.  In essence, the D16 metallicity scale cannot be used to uncover an FMR if it really exists, but the lack of an SFR$-Z$ anti-correlation using D16 metallicities is precisely what is expected if metallicity-SFR fluctuations are driven by fluctuations in the infall rate of pristine gas.

\begin{acknowledgements}

D.K. is supported through the Grant-in-Aid for JSPS Fellows (No. 26-3216). A.R. is grateful to the National Astronomical Observatory of Japan for its support and hospitality while this paper was conceived and set up.  We greatly appreciate the MPA/JHU team for making their catalog public.  Funding for SDSS-III has been provided by the Alfred P. Sloan Foundation, the Participating Institutions, the National Science Foundation, and the US Department of Energy Office of Science. The SDSS-III web site is http://www.sdss3.org/.

\end{acknowledgements}

\end{document}